\documentclass[preprint,cha,onecolumn,superscriptaddress,preprintnumbers,showpacs,amsmath,amssymb]{revtex4-1}
\usepackage{bm}
\usepackage[colorlinks=true,linkcolor=blue,citecolor=blue]{hyperref}
\usepackage{amsthm}
\usepackage{amsfonts}
\usepackage{enumerate}
\usepackage{latexsym}
\usepackage{float}
\usepackage{ifpdf}
\usepackage{graphicx}
\usepackage{makeidx}
\expandafter\ifx\csname package@font\endcsname\relax\else
 \expandafter\expandafter
 \expandafter\usepackage
 \expandafter\expandafter
 \expandafter{\csname package@font\endcsname}
 \fi
\usepackage{color}
\usepackage{times}
\usepackage{multirow}

\begin{document}

\title{Direct evidence of electron-hole compensation for XMR in topologically trivial YBi}

\author{Shaozhu Xiao \footnote[1]{These authors contributed equally to this work.}}
\affiliation{Ningbo Institute of Materials Technology and Engineering, Chinese Academy of Sciences, Ningbo, Zhejiang 315201, China}
\author{Yinxiang Li \footnotemark[1]}
\affiliation{Tin Ka-Ping College of Science, University of Shanghai for Science and Technology, Shanghai 200093, China}
\author{Yong Li \footnotemark[1]}
\affiliation{Institute of Physics, Chinese Academy of Sciences, Beijing 100190, China}
\author{Xiufu Yang \footnotemark[1]}
\affiliation{Ningbo Institute of Materials Technology and Engineering, Chinese Academy of Sciences, Ningbo, Zhejiang 315201, China}
\affiliation{University of Chinese Academy of Sciences, Beijing 100049, China}

\author{Shiju Zhang}
\affiliation{Ningbo Institute of Materials Technology and Engineering, Chinese Academy of Sciences, Ningbo, Zhejiang 315201, China}
\author{Wei Liu}
\affiliation{Ningbo Institute of Materials Technology and Engineering, Chinese Academy of Sciences, Ningbo, Zhejiang 315201, China}
\author{Xianxin Wu}
\affiliation{Max-Planck-Institut für Festkörperforschung, Heisenbergstrasse 1, D-70569 Stuttgart, Germany}
\author{Bin Li}
\affiliation{New Energy Technology Engineering Laboratory of Jiangsu Province and School of Science, Nanjing University of Posts and Telecommunications, Nanjing 210023, China}
\affiliation{National Laboratory of Solid State Microstructures, Nanjing University, Nanjing 210093, China}
\author{Masashi Arita}
\affiliation{Hiroshima Synchrotron Radiation Center, Hiroshima University, Higashi-Hiroshima, Hiroshima 739-0046, Japan}
\author{Kenya Shimada}
\affiliation{Hiroshima Synchrotron Radiation Center, Hiroshima University, Higashi-Hiroshima, Hiroshima 739-0046, Japan}
\author{Youguo Shi}
\email{ygshi@iphy.ac.cn}
\affiliation{Institute of Physics, Chinese Academy of Sciences, Beijing 100190, China}
\author{Shaolong He}
\email{shaolonghe@nimte.ac.cn}
\affiliation{Ningbo Institute of Materials Technology and Engineering, Chinese Academy of Sciences, Ningbo, Zhejiang 315201, China}
\affiliation{Center of Materials Science and Optoelectronics Engineering, University of Chinese Academy of Sciences, 100049 Beijing, China}


\begin{abstract}
The prediction of topological states in rare earth monopnictide compounds has attracted renewed interest. Extreme magnetoresistance (XMR) has also been observed in several nonmagnetic rare earth monopnictide compounds. The origin of XMR in these compounds could be attributed to several mechanisms, such as topologically nontrivial electronic structures and electron-hole carrier balance. YBi is a typical rare earth monopnictide exhibiting XMR, and expected to have a nontrivial electronic structure. In this work, we performed a direct investigation of the electronic structure of YBi by combining angle resolved photoemission spectroscopy and theoretical calculations. Our results show that YBi is topologically trivial without the expected band inversion, and they rule out the topological effect as the cause of XMR in YBi. Furthermore, we directly observed perfect electron-hole compensation in the electronic structure of YBi, which could be the primary mechanism accounting for the XMR.

\end{abstract}

\maketitle
\newpage

\section{INTRODUCTION}

Magnetoresistance (MR) is the effect of a material changing its electrical resistance when an external magnetic field is applied to it. MR in typical nonmagnetic material usually increases quadratically with the magnetic field in a low field, and saturates in a high field. In recent decade, intensive researches of topological materials has led to the discovery of numerous nonmagnetic semimetals that exhibit very large nonsaturating MR\cite{Nature-2014-Ali,PRL-2014-Pletikosic,NC-2017-Kumar,PRB-2016-Chen,Science-2015-Xiong,NP-2015-Shekhar,NM-2015-Liang,PRX-2015-Huang,NP-2016-Tafti,PNAS-2016-Tafti,PRL-2016-Zeng,PRL-2016-He,PRB-2016-Kumar,PNAS-2018-Leahy}, known as extreme magnetoresistance (XMR), which has stimulated significant research interest in its underlying physical mechanisms and potential applications in electronics. While multiple mechanisms have been proposed, the origin of the XMR is still under debate. In some topological semimetals, e.g., Na$_3$Bi\cite{Science-2015-Xiong} and Cd$_3$As$_2$\cite{NM-2015-Liang}, XMR exhibiting linear field dependence is attributed to their topologically nontrivial electronic structures. On the other hand, in some Type-\uppercase\expandafter{\romannumeral2} Weyl semimetals\cite{Nature-2014-Ali,PRL-2014-Pletikosic,NC-2017-Kumar,PRB-2016-Chen} and topologically trivial semimetals\cite{PRL-2016-Zeng,PRL-2016-He}, the compensation between electron and hole charge carriers (carrier compensation) is reported to play an important role in the origin of XMR with quadratic field dependence. Many other mechanisms have also been proposed, such as field-induced change of the Fermi surface\cite{SC-2014-Wang}, field-induced metal-insulator transition\cite{PRB-2016-Li}, and open-orbit Fermi surface topology\cite{PRB-2017-LouR,PRL-2015-Jiang,PRL-2015-Wu}.

Rare earth monopnictides, i.e., RPn (R = rare earth element Sc, Y, La, Ce ...; Pn = group \uppercase\expandafter{\romannumeral6} element N, P, As, Sb, Bi) with rock salt crystal structure, also exhibit quadratic field-dependent XMR behavior\cite{NP-2016-Tafti,PNAS-2016-Tafti,PRL-2016-Zeng,PRL-2016-He,PRB-2016-Kumar,PRB-2018-Pavlosiuk,JMCC-2018-Qian,PRB-2019-Xu,SR-2016-Pavlosiuk,PRB-2017-Wu,PRB-2018-Wang,PRB-2019-Wu,PRB-2017-Yang}. The topological properties of the RPn family are interesting, and the origin of XMR in RPn is multiplex. A previous theoretical work\cite{arXiv-2015-Zeng}, which rekindled significant research interest in RPn, predicted that lanthanum monopnictides, LaPn (Pn = N, P, As, Sb, Bi),  could be topological semimetals or topological insulators as a consequence of band inversion around the X point of the bulk fcc Brillouin zone. However, later studies\cite{PRL-2016-Zeng, PRB-2016-Guo} indicated that the band inversion was often overestimated in the calculations of the LaPn family, and only LaBi is topologically nontrivial among them. Because angle resolved photoemission spectroscopy (ARPES) has the ability to directly reveal the band structures of materials, the band topologies of numerous RPn compounds have been experimentally studied\cite{PRL-2016-He,PRL-2016-Zeng,PRB-2016-Kumar,PRB-2018-Wang,PRB-2019-Wu,PRB-2017-Yang,PRB-2016-Niu,PRB-2017-Lou,NC-2017-Nayak,PRM-2018-Jiang,PRB-2017-Oinuma,npjQM-2018-Nummy,PRB-2018-Li-CeBi,PRL-2018-Kuroda,PRB-2019-Li-CeBi}. Topological band inversion and nontrivial surface states were observed in compounds with group \uppercase\expandafter{\romannumeral6} element Bi, such as LaBi\cite{PRB-2016-Kumar,PRB-2016-Niu,PRB-2017-Lou,NC-2017-Nayak,PRM-2018-Jiang}, SmBi\cite{PRB-2019-Wu,PRB-2018-Li-CeBi}, and CeBi\cite{PRB-2018-Li-CeBi,PRL-2018-Kuroda,PRB-2019-Li-CeBi}, while compounds with other group \uppercase\expandafter{\romannumeral6} elements were confirmed to be topologically trivial without band inversion\cite{PRL-2016-He,PRL-2016-Zeng,PRB-2018-Wang,PRB-2019-Wu,PRB-2017-Yang,PRB-2017-Oinuma,npjQM-2018-Nummy,PRL-2018-Kuroda}. In those topologically trivial compounds, XMR behaviours are widely attributed to the classical electron-hole carrier compensation scenario\cite{PRL-2016-He,PRL-2016-Zeng}.

Yttrium monopnictide YBi, belonging to the RPn family, has been reported to also exhibit XMR effect\cite{PRB-2018-Pavlosiuk,JMCC-2018-Qian,PRB-2019-Xu}. Magnetotransport investigations and first-principle calculations attribute the XMR in YBi mainly to the electron-hole compensation\cite{PRB-2018-Pavlosiuk,JMCC-2018-Qian,PRB-2019-Xu}. However, there is no direct measurement of its electronic structure to confirm electron-hole compensation in YBi. Moreover, one can not rule out the presence of the role of topology in the origin of XMR in YBi because several calculations have predicted that YBi is topologically nontrivial\cite{PRB-2018-Pavlosiuk,JMCC-2018-Qian,PRB-2019-Xu}. In this work, we have employed ARPES to directly probe the band structure of YBi. There is no band inversions observed in the ARPES results. The combination of an ARPES experiment and first-principle calculation directly revealed that the electron and hole carriers in YBi are nearly compensated, which could be responsible for the XMR in topologically trivial YBi.

\section{EXPERIMENTAL}

Single crystals of YBi were grown out of In flux. High-purity materials Y, Bi, and In were mixed in a molar ratio of Y : Bi : In = 1 : 1 : 20. And they were loaded into an alumina crucible in a highly-purified argon atmosphere glove box. The crucible was sealed in an evacuated quartz tube, heated to 1373 K and then cooled down to 973 K at a rate of 2 K/h. The crystals were separated from the In flux by centrifugation.

ARPES measurements were performed on beamline 9A \cite{NIMA-Matsui-BL9A, SRL-Arita-BL9A} of the Hiroshima Synchrotron Radiation Center (HiSOR), with a total energy resolution of $\sim$ 25 meV and a base pressure better than 5 $\times$ 10$^{-9}$ Pa. The YBi samples for ARPES measurements were prepared in a highly-purified argon atmosphere glove-box to avoid degradation of YBi during the curing process of silver glue. Fresh surfaces for ARPES measurements were obtained by cleaving the samples \textit{in situ} along the (001) plane at approximately 20 K. During ARPES measurements, the sample temperature was kept at approximately 20 K.

The band structure calculations were performed based on the density-functional theory (DFT), using an all-electron full potential WIEN2K code \cite{Blaha}. The exchange correlation energy was treated using Perdew-Burke-Ernzerhof(PBE)\cite{PRL-1996-Perdew} and modified Becke-Johnson (mBJ)\cite{PRL-2009-Tran} functionals. The spin-orbit coupling (SOC) effect was included in all calculations. The mBJ method could overcome the band gap underestimation and band inversion overestimation using the PBE method\cite{PRL-2016-He,PRB-2016-Guo,PRL-2016-Zeng,PRB-2017-Oinuma,SR-2018-Dey,CP-2018-Duan-LnPn,PRB-2018-Li-CeBi,PRB-2019-Li-CeBi}, which was also verified by our experimental results. Therefore, unless specifically indicated, the calculation results in this work were obtained using the mBJ method. During the calculations, the plane-wave cutoff parameter was selected as $R_{MT}\times K_{max}$ = 7 (product of muffin-tin radius vector and wave vector in a reciprocal lattice within the Brillouin zone). The spherical harmonic inside the muffin-tin spheres was $l_{max}$ = 10 and the Fourier expanded charge density was $G_{max}$ = 12. A set of 10$\times$10$\times$10 special k-points was used for integration over the Brillouin zone of a unit cell.

\section{RESULTS AND DISCUSSION}

\begin{figure*}[]
\includegraphics[width=0.6\textwidth]{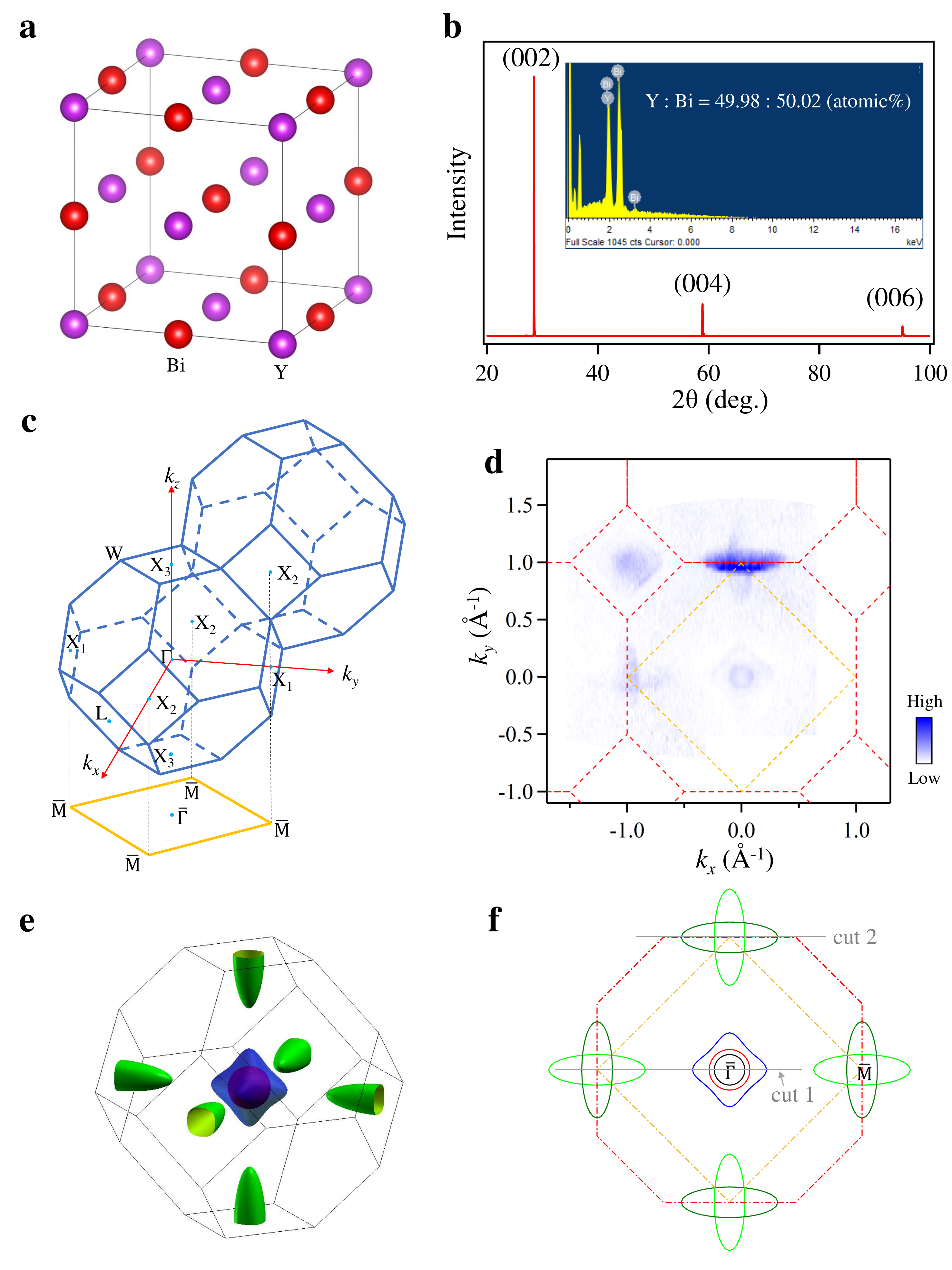}
\caption{\label{Fig1}(a) Schematic crystal structure of YBi. (b) XRD $\theta-2\theta$  pattern on the cleaved surface of YBi after ARPES measurements. Inset: energy dispersive x-ray (EDX) spectrum of YBi. (c) Three-dimensional (3D) bulk Brillouin zone of YBi and the projected (001) surface Brillouin zone. Three inequivalent bulk X points are labelled as X$_1$, X$_2$ and X$_3$. X$_1$ and X$_2$ are projected onto the surface $\overline{\rm M}$ point while $\Gamma$ and X$_3$ are projected onto the surface $\overline{\Gamma}$ point. (d) Fermi surface of YBi measured by ARPES using photon energy $h\nu$ = 25 eV. Yellow dashed lines represent the 2D surface Brillouin zone. (e) Calculated 3D Fermi surface of YBi in the first Brillouin zone. (f) Projected Fermi surface on the (001) surface obtained by calculation. Yellow dashed lines represent the surface Brillouin zone. The three hole pockets around $\overline{\Gamma}$ point are projected from $\Gamma$ point (red and blue pockets) and X$_3$ point (black pocket), and the two electron pockets around $\overline{\rm M}$ point are projected from X$_1$ points (light green pocket) and X$_2$ points (dark green pocket). }
\end{figure*}

YBi crystallizes in a NaCl-type face-centered cubic (FCC) structure (space group $Fm\bar{3}m$), as illustrated in Fig.~\ref{Fig1}(a). The x-ray diffraction (XRD) $\theta-2\theta$ pattern on the cleaved surface of YBi after ARPES measurements is shown in Fig.~\ref{Fig1}(b). Only $(00l)$ peaks of YBi were observed without any impurity derived ones, confirming the crystalline quality of YBi and that the cleaved surface was the (001) surface. The stoichiometry of the YBi sample was also verified by energy dispersive x-ray spectroscopy (EDX), as shown in the inset of Fig.~\ref{Fig1}(b).

Fig.~\ref{Fig1}(c) plots the three-dimensional (3D) bulk Brillouin zone and the projected (001) surface Brillouin zone of YBi. Considering the projection onto the (001) surface Brillouin zone, there are three inequivalent X points in the 3D bulk Brillouin zone. The X$_1$ and X$_2$ points are projected onto the surface $\overline{\rm M}$ point, while the $\Gamma$ and X$_3$ points are projected onto the surface $\overline{\Gamma}$ point. Figs.~\ref{Fig1}(d)-\ref{Fig1}(f) show the Fermi surface of YBi measured by ARPES, the calculated 3D bulk Fermi surface and its projection on the (001) surface Brillouin zone, respectively. As shown in Fig.~\ref{Fig1}(e), the bulk Fermi surface of YBi consists of two hole pockets (red inside and blue outside) at the $\Gamma$ point and one electron pocket (green) at each X point, which is a common feature in RPn materials, e.g., LaSb, YSb, and LaBi\cite{PNAS-2016-Tafti,PRL-2016-Zeng,PRL-2016-He,PRB-2016-Kumar}. We now consider the ARPES measured Fermi surface of YBi in Fig.~\ref{Fig1}(d), which agrees well with the projected Fermi surface by calculation in Fig.~\ref{Fig1}(f). Around the surface $\overline{\Gamma}$ point, there are three hole-like pockets, where the outer two (red and blue) are projected from the $\Gamma$ point and the inner black one from the X$_3$ point. Around the surface $\overline{\rm M}$ point, there are two intersecting ellipse-shaped electron-like pockets. The one (light green) with its long-axis along the $\overline{\Gamma}$-$\overline{\rm M}$ direction is projected from the X$_1$ point and the other one (dark green) from the X$_2$ point.

\begin{figure*}[]
\includegraphics[width=\textwidth]{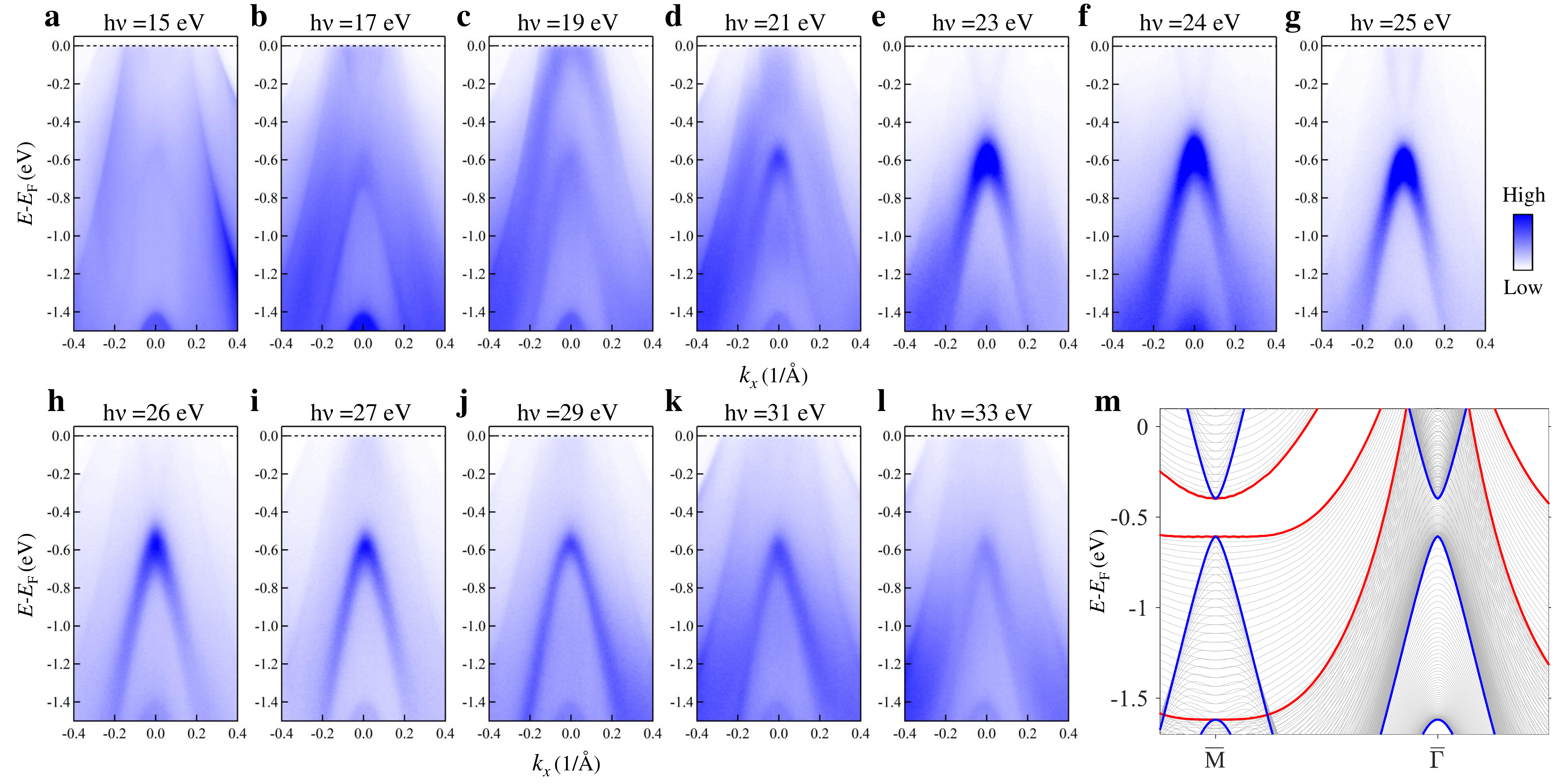}
\caption{\label{Fig2}(a)-(l) Band dispersions of YBi along $\overline{\Gamma}$-$\overline{\rm M}$ direction measured by ARPES with various photon energies as labelled. (m) Projected bulk band structures of YBi on the (001) surface obtained by calculation. The red (blue) curves indicate the band dispersions on $k_z = 0$ ($k_z = \pi$) plane.}
\end{figure*}

Fermi surface features of YBi at different $k_z$ coexist in one ARPES measured Fermi surface, as discussed above and shown in Fig.~\ref{Fig1}. This can be attributed to the $k_z$ broadening effect, which is common in the RPn compounds\cite{PRB-2018-Wang,PRB-2019-Wu,PRB-2016-Niu,PRM-2018-Jiang,PRB-2017-Oinuma,PRB-2018-Li-CeBi}. As a result of short escape depth of photoelectrons excited by the ultraviolet photons (e.g., $h\nu = 15 \sim 40$ eV in this work), i.e. $\Delta x <\sim 5 \textrm{\AA}$\cite{SIA-1979-Seah}, corresponding uncertainty of $k_z$ becomes larger than $\sim 2\pi/\Delta x \sim 1~\textrm{\AA}^{-1}$, which is nearly the distance between $\Gamma$ and X points, $1~\textrm{\AA}^{-1}$. To further verify this, we conducted photon energy-dependent ARPES measurements on YBi. The band dispersions of YBi along the $\overline{\Gamma}$-$\overline{\rm M}$ direction obtained with various photon energies are shown in Figs.~\ref{Fig2}(a)-\ref{Fig2}(l). Except the spectral weight, the band dispersions show little change with photon energy. Fig.~\ref{Fig2}(m) shows the projected bulk band structure of YBi on the (001) surface obtained from the calculations, comprising bulk band dispersions along $\overline{\Gamma}$-$\overline{\rm M}$ over the entire $k_z$ range. The red curves indicate the band dispersions on $k_z = 0$, and the blue curves indicate the band dispersions on $k_z = \pi$. The band dispersions from other $k_z$ fill the space between those from $k_z = 0$ and $\pi$, as shown in grey.

While the ARPES spectral intensity is clearly modulated depending on the incident photon energies as shown in Figs.~\ref{Fig2}(a)-\ref{Fig2}(l), we cannot see a systematic $k_z$ dependent energy shift due to the $k_z$ broadening effect. One can see, however, $k_z$ weighted spectral features, which are not contradictory to the theoretical $k_z$ projected results as shown in Fig.~\ref{Fig2}(m). In the ARPES image obtained using $h\nu = 25$ eV, as shown in Fig.~\ref{Fig2}(g), the major features with the largest spectral weights are two cone-like structures at the $\overline{\Gamma}$ point, which is consistent with the blue curves in Fig.~\ref{Fig2}(m). This suggests that, if considering the photon energy vs $k_z$ correspondence, a photon energy of 25 eV is close to $k_z = \pi$. On the one hand, the $k_z$ broadening effect mixes the bulk state and surface state features during the photon-energy dependent ARPES experiments, making the data analyses much more complicated. On the other hand, the sizeable $k_z$ broadening collects bulk state features from a broad $k_z$ range in one fixed ARPES photon energy measurement, making some ARPES analyses feasible without changing the photon energy. We assume that the ARPES spectral features for the $k_z$ corresponding to the high symmetry points should have stronger intensity because the energy bands should have the stationary points at these $k_z$ points. Thus, in the following, we use the ARPES spectra obtained with a fixed photon energy to analyse the band topology of YBi.

\begin{figure*}[]
\includegraphics[width=0.8\textwidth]{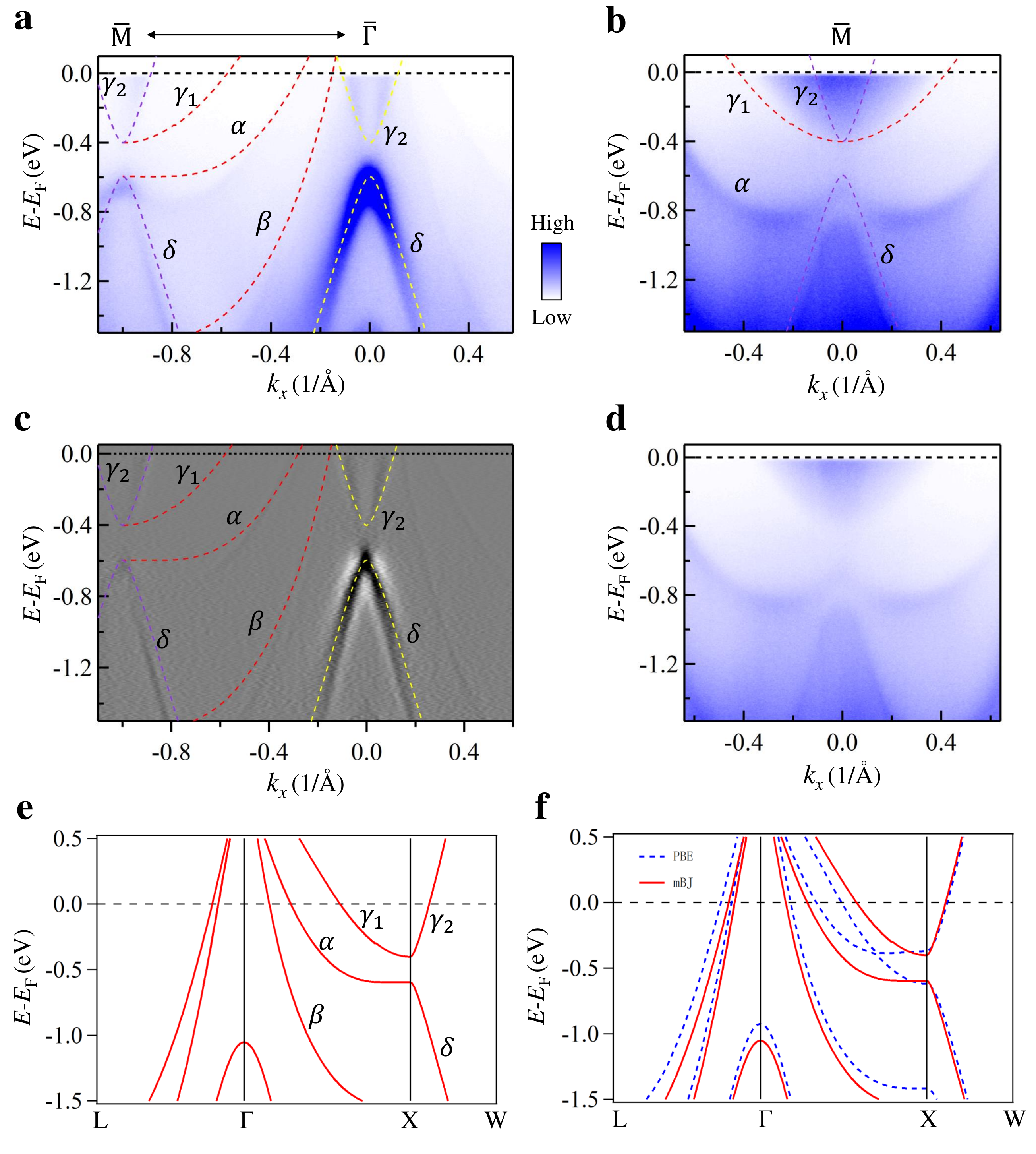}
\caption{\label{Fig3} (a) Band structures of YBi along $\overline{\rm M}$-$\overline{\Gamma}$-$\overline{\rm M}$ direction [cut 1 grey line in Fig.~\ref{Fig1}(f)] measured by ARPES with photon energy of 25 eV. The dashed curves are the calculated bands. (b) Same as (a), but along $\overline{\Gamma}$-$\overline{\rm M}$-$\overline{\Gamma}$ [cut 2 grey line in Fig.~\ref{Fig1}(f)]. (c) Second derivative image of (a) with respect to the momentum. (d) Same as (b), but without superimposing the calculated bands in order not to cover the details of the ARPES spectrum. (e) The band structures of YBi along high-symmetry directions calculated by mBJ method. (f) Comparison of the band structures of YBi calculated by mBJ method and PBE method.}
\end{figure*}

Figs.~\ref{Fig3}(a) and \ref{Fig3}(b) show  the band structures of YBi measured by ARPES along the high-symmetry directions of $\overline{\rm M}$-$\overline{\Gamma}$-$\overline{\rm M}$ [cut 1 grey line in Fig.~\ref{Fig1}(f)] and $\overline{\Gamma}$-$\overline{\rm M}$-$\overline{\Gamma}$ [cut 2 grey line in Fig.~\ref{Fig1}(f)], respectively, where the calculated band structures are superimposed on top of them. For clarity, the calculated bulk band structures of YBi along high-symmetry lines are shown in Fig.~\ref{Fig3}(e). The ARPES spectrum shown in Fig.~\ref{Fig3}(a) agrees well with the calculation, especially the projected bulk band structures shown in Fig.~\ref{Fig2}(m). The good agreement between the ARPES measurement and the calculation can also be clearly seen from the second derivative image of Fig.~\ref{Fig3}(a), as shown in Fig.~\ref{Fig3}(c). As discussed earlier, the $k_z$ broadening effect in YBi allows ARPES measurements to cover the band structures across the entire $k_z$ range even with one photon energy. Therefore, the bands indicated by the red dashed curves of $\alpha$, $\beta$, and $\gamma_1$ are likely projected from the bulk bands along $\Gamma$-$X_1$ on the $k_z = 0$ plane, and the bands indicated by the yellow (purple) dashed curves of $\gamma_2$ and $\delta$ are projected from the bulk bands along $W$-$X_3$-$W$ ($W$-$X_2$-$W$) on the $k_z = \pi$ plane. Similarly, for the ARPES spectrum shown in Fig.~\ref{Fig3}(b), $\gamma_2$ and $\delta$ are projected from the bulk bands along $W$-$X_2$-$W$ on the $k_z = 0$ plane, and $\gamma_1$ and $\alpha$ are projected from those along $\Gamma$-$X_1$ in the second bulk Brillouin zone. For RPn compounds, the DFT calculations using the PBE functional usually result in the overestimation of band anti-crossing, and result in a misleading prediction of nontrivial topology\cite{PRL-2016-He,PRB-2016-Guo,PRL-2016-Zeng,PRB-2017-Oinuma,SR-2018-Dey,CP-2018-Duan-LnPn,PRB-2018-Li-CeBi,PRB-2019-Li-CeBi}, which is also the case for YBi\cite{PRB-2018-Pavlosiuk,JMCC-2018-Qian,PRB-2019-Xu}, as shown in Fig.~\ref{Fig3}(f). In contrast, calculations using the mBJ functional agree well with the experimental results. Thus, by combining the ARPES experimental and  theoretical calculation results, one can conclude that no anti-crossing of bands and no band inversion exist along the bulk $\Gamma$-$X$ direction.

\begin{table*}[]
\caption{\label{table1}Enclosed volumes of Fermi surface pockets and corresponding carrier densities in YBi.}
\begin{tabular}{cccccc}
\hline
\multirow{2}{*}{~~Pocket~~} & \multirow{2}{*}{~~Type~~} & \multicolumn{2}{c}{~~Theoretical calculation~~} & \multicolumn{2}{c}{\begin{tabular}[c]{@{}c@{}}Theoretical calculation \\ modified by ARPES measurement\end{tabular}} \\ \cline{3-6}
               &              & \begin{tabular}[c]{@{}c@{}}~~Volume \\ ({\AA}$^{-3}$) \end{tabular} & \begin{tabular}[c]{@{}c@{}} Carrier density \\ ($\times 10^{20}$~cm$^{-3}$)  \end{tabular} &  \begin{tabular}[c]{@{}c@{}}~~~~~Volume \\ ({\AA}$^{-3}$) \end{tabular} &\begin{tabular}[c]{@{}c@{}} Carrier density \\ ($\times 10^{20}$~cm$^{-3}$)  \end{tabular}        \\ \hline
$\alpha$                       & hole                  & 0.0479      & 3.86      & 0.0445        & 3.59           \\ \hline
$\beta$                        & hole                  & 0.0165      & 1.33      & 0.0156        & 1.25           \\ \hline
$\gamma$\footnotemark[1]       & electron              & 0.0647      & 5.22      & 0.0597        & 4.82           \\ \hline
\end{tabular}
\footnotetext[1]{Note: There are three $\gamma$ pockets in one Brillouin zone, so the volume and corresponding carrier density of the $\gamma$ pocket was multiplied by 3.}
\end{table*}

As the band inversion is absent, YBi is topologically trivial, and hence the XMR of YBi could be a result of electron-hole compensation, which resembles that in LaSb\cite{PRL-2016-Zeng} and YSb\cite{PRL-2016-He}. To verify this, we calculated the electron and hole carrier densities. For convenience, we hereafter label the two hole pockets located around the $\Gamma$ point [see Fig.~\ref{Fig1}(e)] as $\alpha$ and $\beta$ (for outer and inner, respetively), and the electron pocket located around the $X$ point as $\gamma$. The carrier densities were calculated based on the calculation of the enclosed volumes of the corresponding Fermi surface pockets, and the results are shown in Table \ref{table1}. From pure theoretical calculation, the enclosed volumes of the $\alpha$, $\beta$, and $\gamma$ Fermi surface pockets in one Brillouin zone are 0.0479, 0.0165, and 0.0647 {\AA}$^{-3}$, corresponding to carrier densities of $3.86 \times 10^{20}$, $1.33 \times 10^{20}$, and $5.22 \times 10^{20}$ cm$^{-3}$, respectively, suggesting a perfect electron-hole compensation. In order to improve the carrier density calculation accuracy, we modified the theoretical calculations using the ARPES measured Fermi momenta at high-symmetry lines. The corrected volumes of $\alpha$, $\beta$, and $\gamma$ are 0.0445, 0.0156, and 0.0597 {\AA}$^{-3}$, corresponding to carrier densities of $3.59 \times 10^{20}$, $1.25 \times 10^{20}$, and $4.82 \times 10^{20}$ cm$^{-3}$, respectively, and the ratio of electron-to-hole carrier density is $n_e/n_h = 0.996$. Therefore, YBi is a compensated semimetal with nearly perfect electron-hole carrier balance, which could be the origin of the XMR in YBi.

\begin{figure*}[]
\includegraphics[width=\textwidth]{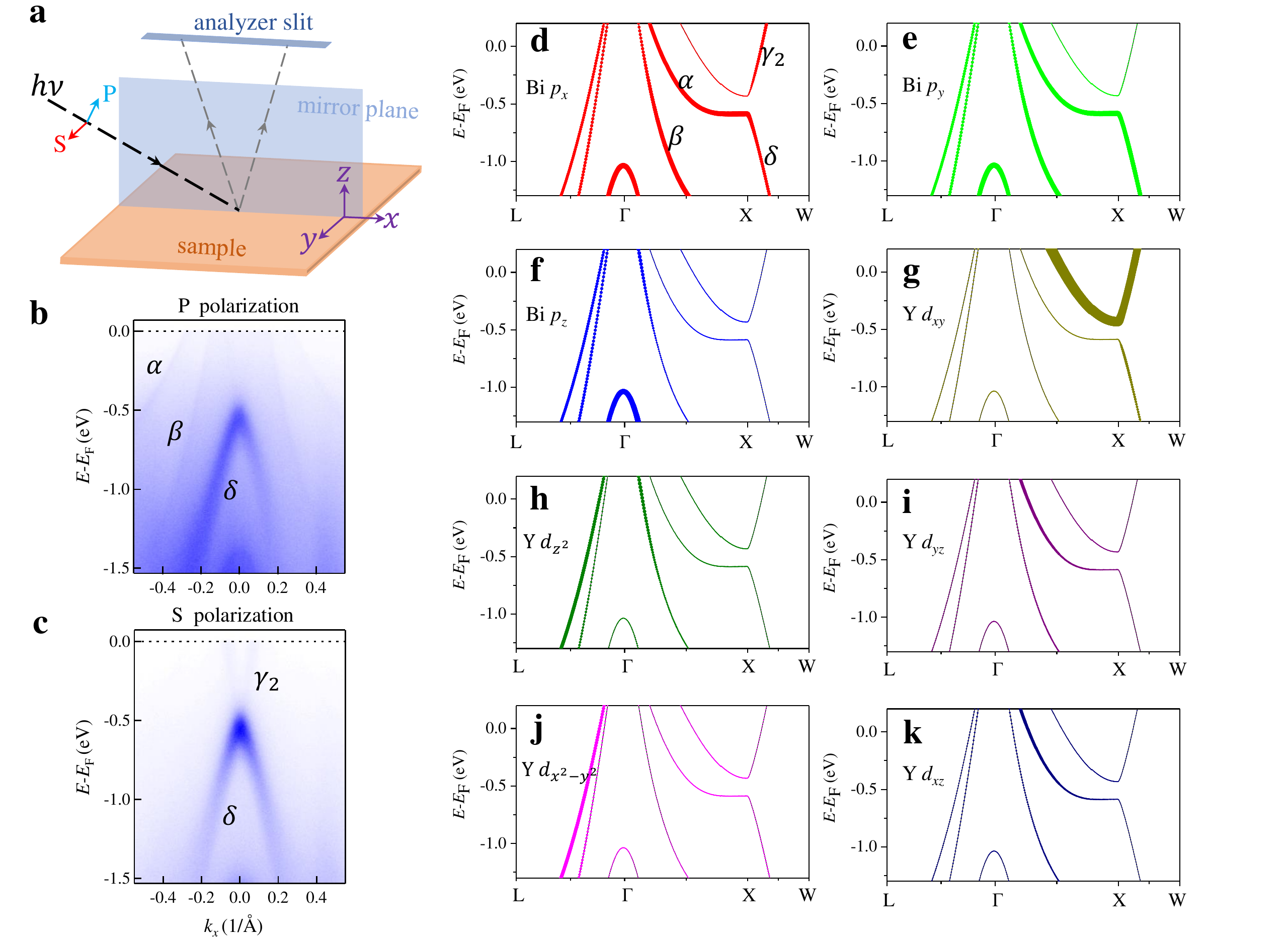}
\caption{\label{Fig4}(a) Experimental setup of the polarization-dependent ARPES measurement. The electric field direction of the $p$ (or $s$) incident photons is parallel (or perpendicular) to the mirror plane defined by the analyzer slit and the sample surface normal, that is, the $xz$ plane. (b) Band structures of YBi along $\overline{\rm M}$-$\overline{\Gamma}$-$\overline{\rm M}$ measured by $p$-polarization photons. (c) Same as (b), but measured by $s$-polarization photons. (d)-(k) Calculated band structures of YBi with projections on different orbitals as labelled. The thickness of the curves represents the weight (or the contribution) of the corresponding orbital.}
\end{figure*}

Finally, to further study the origin of the bands $\gamma_2$ and $\delta$ and further support the previous discussion that they are projected from the bulk bands along $W$-$X$-$W$, we conducted polarization-dependent ARPES experiments on YBi. Fig.~\ref{Fig4}(a) shows the experimental setup of the polarization-dependent ARPES. Figs.~\ref{Fig4}(b) and \ref{Fig4}(c) show the band structures of YBi along $\overline{\rm M}$-$\overline{\Gamma}$-$\overline{\rm M}$ measured by $p$-polarization and $s$-polarization photons, respectively. Based on the matrix element effect of ARPES, the even orbitals, which are even with respect to the mirror plane ($xz$ plane), could only be probed by $p$-polarization photons, while the odd orbitals could only be probed by $s$-polarization photons. Therefore, the bands with contributions from the Bi orbitals $p_x$ and $p_z$ and the Y orbitals $d_{z^2}$, $d_{x^2-y^2}$, and $d_{xz}$ could only be observed by $p$-polarization photons, while the bands with contributions from the Bi orbital $p_y$ and Y orbitals $d_{xy}$ and $d_{yz}$ could only be observed by $s$-polarization photons. As shown in Figs.~\ref{Fig4}(d)-\ref{Fig4}(k), the $\gamma_2$ band is mainly contributed by the odd orbital Y $d_{xy}$, and could mainly be probed by the $s$-polarization photons. In contrast, the $\delta$ band is contributed by both even and odd orbitals, and could be probed by both types of polarization. These results are consistent with the ARPES spectra shown in Figs.~\ref{Fig4}(b) and \ref{Fig4}(c), supporting that the bands $\gamma_2$ and $\delta$ are projected from the bulk bands along $W$-$X$-$W$ and are not the features of surface states.

\section{CONCLUSION}
In summary, the electronic structures of YBi were comprehensively investigated by means of ARPES in combination with the theoretical DFT calculations. The $k_z$ broadening effect in YBi is sizeable, and it collects bulk band features across over the entire $k_z$ range in a fixed-photon-energy ARPES spectra. Combining the ARPES experimental and theoretical calculation results, we find that band inversions are absent in YBi, leading to a topologically trivial electronic structure. The electron and hole carrier densities in YBi are almost perfectly balanced, indicating that YBi is a compensated semimetal, which could be the origin of the XMR in YBi.

\section{ACKNOWLEDGMENTS}
This work is supported by the National Key Research and Development Program of China (No. 2016YFA0300600, No. 2017YFA0303600, and 2017YFA0302901 ), the National Natural Science Foundation of China (Grant Nos. 11674367,11974364, U2032207, and U2032204), the Natural Science Foundation of Zhejiang, China (Grant No. LZ18A040002), the Ningbo Science and Technology Bureau (Grant No. 2018B10060). S.L.H. would like also to acknowledge the Ningbo 3315 program. Y.G.S. would like also to acknowledge the Strategic Priority Research Program (B) of the Chinese Academy of Sciences (No. XDB33000000), the K. C. Wong Education Foundation (GJTD-2018-01). We thank the Hiroshima Synchrotron Radiation Center for providing beamtime ( project No. 19BG025).

\textbf{Author contributions}

S.L.H. and S.Z.X conceived the project. Y.L. and Y.G.S. grew the samples. S.Z.X., X.F.Y., S.J. Z., and W.L. carried out the ARPES measurements. Y.X.L., X.X.W. and B.L. performed the theoretical calculations. S.L.H. and S.Z.X. analyzed the data and wrote the paper. All authors discussed the results and commented on the manuscript.

\end{document}